\documentclass[review,authoryear,12pt]{elsarticle}

\usepackage{natbib}

\begin{document}

\begin{frontmatter}

\title{Are Large Trojan Asteroids Salty? An Observational, Theoretical, and Experimental Study} 


\author
{Bin Yang$^{1}$, Paul Lucey$^{2}$, Timothy Glotch$^{3}$\\
\normalsize{$^{1}$Institute for Astronomy, University of Hawaii,}\\
\normalsize{2680 Woodlawn Dr., Honolulu, HI 96822, USA}\\
\normalsize{$^{2}$Hawaii Institute of Geophysics and Planetology,
University of Hawaii, }\\
\normalsize{1680 East West Road, Honolulu, HI 96822, USA}\\
\normalsize{$^{3}$Department of Geosciences,Stony Brook University,}\\
\normalsize{255 Earth and Space Science, Stony Brook, NY 11794-2100, USA}\\
}

\begin{abstract}

With a total mass similar to the main asteroid belt, the Jovian Trojan asteroids are a major feature in the solar system. Based upon the thermal infrared spectra of the largest Trojans obtained with the Spitzer space telescope, \citet{emery:2006} suggested that the surfaces of these Trojans may consist of fine-grained silicates suspended in a transparent matrix.

To explore the transparent matrix hypothesis, we adopted a modified radiative transfer model to fit the Trojan spectra simultaneously both in the near and the thermal infrared regions. Our model shows that the Trojan spectra over a wide wavelength range can be consistently explained by fine grained silicates (1-5 wt.\%) and highly absorbing material (e.g.\ carbon or iron, 2-10 wt.\%) suspended in a transparent matrix. The matrix is consistent with a deposit of salt on the surfaces of the large Trojans. However, this consistency is not an actual detection of salt and other alternatives may still be possible. We suggest that early in the Solar System history, short-lived radionuclides heated ice-rich Trojans and caused melting, internal circulation of water and dissolution of soluble materials. Briny water volcanism were facilitated by internal volatiles and a possibly global sill of frozen brine was formed beneath the cold primitive crust. The frozen brine layer was likely to be evacuated by impact erosions and evaporation of the exposed brines eventually left a lag deposit of salt. Over the Solar System's history, fine dust from comets or impacts contaminated and colored these salty surfaces of the Trojans to produce the spectral properties observed today.

\end{abstract}


\end{frontmatter}


\section{Introduction}

Along Jupiter$'$s orbit there are two regions of gravitational stability occupied by swarms of small bodies referred to as the Trojan asteroids.  The Trojans larger than 1 km are comparable in number to the asteroids between Mars and Jupiter \citep{jewitt:2000, yoshida:2005} and hence are a major feature in the solar system. Their nature is of high interest owing to their intermediate position between the main belt asteroids and the Kuiper belt population beyond Neptune. The origin of the Trojan population is currently unknown; they may have formed in Jupiter$'$s vicinity, or elsewhere in the outer solar system and were captured subsequently by Jupiter \citep{marzari:1998,fleming:2000,morbidelli:2005}.  The composition of these objects holds clues both to the nature of their origin, and the evolution of the early solar system.

Spectroscopic measurements of the large to medium-sized Trojans, using ground-based telescopes, have yielded high quality spectra in the visible and the near infrared \citep[NIR;][]{cruikshank:2001, emery:2003, dotto:2006, fornasier:2007, emery:2011, yang:2011}. The albedos of Trojans are low, which is about 4\% \citep{fernandez:2003, grav:2011}. Their spectra in between 0.4 to 2.5 $\mu$m tend to be relatively featureless and exhibit red spectral slopes; that is the reflection increases steadily from shorter wavelengths (visible) to longer wavelengths (infrared). Previous radiative transfer models of Trojans require dark components (e.g.\ carbon or organics) to explain their low albedos and other components (e.g.\ silicates) to fine-tune the model results \citep{cruikshank:2001, emery:2004}. 

Recent thermal infrared (TIR, 6-30 $\mu$m) observations of the four largest Trojans using the Spitzer space telescope provide an important constraint on the surface composition of the Trojans \citep{emery:2006, mueller:2010}. Despite the lack of any evidence of comet-like activity for Trojans, such as the presence of coma or tail, Emery et al.\ noted that the Trojan spectra most closely resembled the comae of active comets. This finding is quite surprising, because the existing observations suggest that Trojans have asteroidal surfaces, whose scattering and spectral properties 
are distinguishably different from cometary coma (i.\ e.\ diffuse dust clouds of silicates with other components suspended in space).  In an effort to understand the TIR spectra, \citet{emery:2006} conducted an exhaustive spectroscopic mixing analysis of these spectra implementing all methods commonly used in planetary spectroscopic analysis and found no entirely satisfactory solutions. Nevertheless, Emery et al.\ have probably ruled out some special combination of previously suggested components situated in a conventional particulate regolith that can explain these spectra. 

Because cometary comae consist of well-separated dust particles of silicates and carbon (among other species), \citet{emery:2006} proposed that the surfaces of the Trojans might consist of silicate particles entrained in a transparent matrix of unspecified material. Recently \citet{king:2011} have shown that silicate material (particulate ordinary chondrite) suspended in a matrix of salt -- highly transparent in the infrared -- can qualitatively enhance the contrast of the 10 $\mu$m distinguishing characteristics of the Trojan spectra, supporting the suggestion of Emery et al.\ that such a mixing condition can explain their TIR spectra.  However, it has not been demonstrated that the NIR spectra of the Trojans can be explained by a suspension of material in a transparent matrix, with model abundances similar at both wavelengths. 

In this paper we reproduced the NIR spectra of several Trojans computing from a radiative transfer model that includes the effects of extremely fine-grained materials suspended in a transparent matrix based upon the suggestions of \citet{emery:2006} and \citet{king:2011}. Among our findings is the need for strong absorbers (iron or carbon) that could potentially obscure the TIR 10 $\mu$m spectral feature. Using theory and experiment, we examined the impact of the strong absorbers on the TIR emission features. We presented a plausibility argument that salt suggested by \citet{king:2011} could be the candidate for the transparent matrix. We also discussed a possible scenario for the origin of salt on the surface of the large Trojans. 

\section{Observations}
\cite{emery:2011} showed that the Trojans exhibit two spectral types that differ in their spectral slopes.  We obtained spectra of examples of both of the two groups (namely the ``red" type and the ``grey" type) using the NASA Infrared Telescope Facility (IRTF) 3-m telescope covering the wavelength range from 0.8 $\mu$m to 2.5 $\mu$m for all of our observations at a spectral resolution ($\lambda$/$\Delta\lambda$) $\sim$ 130. Details of these observations are provided in \citep{yang:2011}. 

\section{Radiative Transfer Model}
The radiative transfer model used in this paper treats the problem of submicroscopic absorbers and scatterers in a transparent matrix. The model is a slight modification to that of \citet{lucey:2011} and is in turn based on that of \citet{hapke:2001}, which addresses the problem of modeling the spectra of very small absorbing grains in transparent matrices in an effort to explain aspects of the unusual optical properties of the Moon. Hapke's approach is to modify the absorption coefficient of transparent hosts with the effects of submicroscopic particles and embed this in his overall model for bidirectional reflectance \citep{hapke:1981, hapke:1993}.  \citet{hapke:2001} used Maxwell-Garnet theory to derive the absorbing effects of particles with Rayleigh $x$ (2$\pi r / \lambda$, $r$ is particle radius, $\lambda$ is the wavelength, both expressed in the same units) parameter values much less than one. \citet{lucey:2008} used experimental data to show that HapkeÕs treatment worked well for the very small particle sizes, but at values of $x$ above unity, Hapke$'$s treatment departed strongly from experiment. \citet{lucey:2011} adopted the same approach used by \citet{hapke:2001}, but replaced the absorption derived from Maxwell-Garnet theory with size dependent Mie-theory and showed that this model reproduced measured spectra of mixtures over a wider range of sizes than using the Maxwell-Garnet approximation. 

\subsection{Modified Hapke Model}
In this paper, we adopt Lucey and Riner$'$s approach, but also add the effects of scattering by the extremely fine-grained olivine particles. In Hapke$'$s treatments, scattering by small aspersions in grains is described by an empirical scattering parameter $s$ that modifies the single scattering albedo of the grain \citep{hapke:1981, hapke:1993}.  To compute the parameter $s$ using Mie theory, we first calculate the scattering efficiency $q_s$ taking note of the fact that the absorbing particle is immersed in a host medium of complex refractive index $n_h$, so the effective complex refractive index used in the Mie calculations is $n_{eff}$=$n_g$/$n_h$. The effect of the scattering particles is their scattering efficiencies weighted by the cross sectional area of the particles. This is expressed in terms of mass fraction in equation (1) for which a derivation is given in \citet{lucey:2011} (in that case the effective absorption coefficient of absorbers is derived instead of s, but the derivation is identical, replacing $q_a$ with $q_s$).

\begin{equation}
\label{eq1}
s_{Mie}=\frac{3 q_s M_p \rho_h}{d_p \rho_p}
\end{equation}
where $s_{Mie}$ is the scattering parameter, $q_s$ is the scattering efficiency computed from Mie theory, $M_p$ is the mass fraction of the scattering particle, $\rho_p [g/cc]$ and $\rho_h [g/cc]$ are the densities of the scattering particle and medium respectively, and $d_p$ is the size of the scattering particle in centimeter. This assumes a relatively low density of scatterers to avoid inter-particle shadowing, and this low density permits the use of Mie scattering that includes diffraction without correction. Because the scattering efficiency of absorbing particles is so low compared to their absorption efficiency the scattering effects of the absorbing particles (iron or carbon) are neglected. 

\begin{figure}[t]
\begin{center}
\vspace{-1.8 cm}
\vspace{-0.3 cm}\includegraphics[width=3in,angle=0]{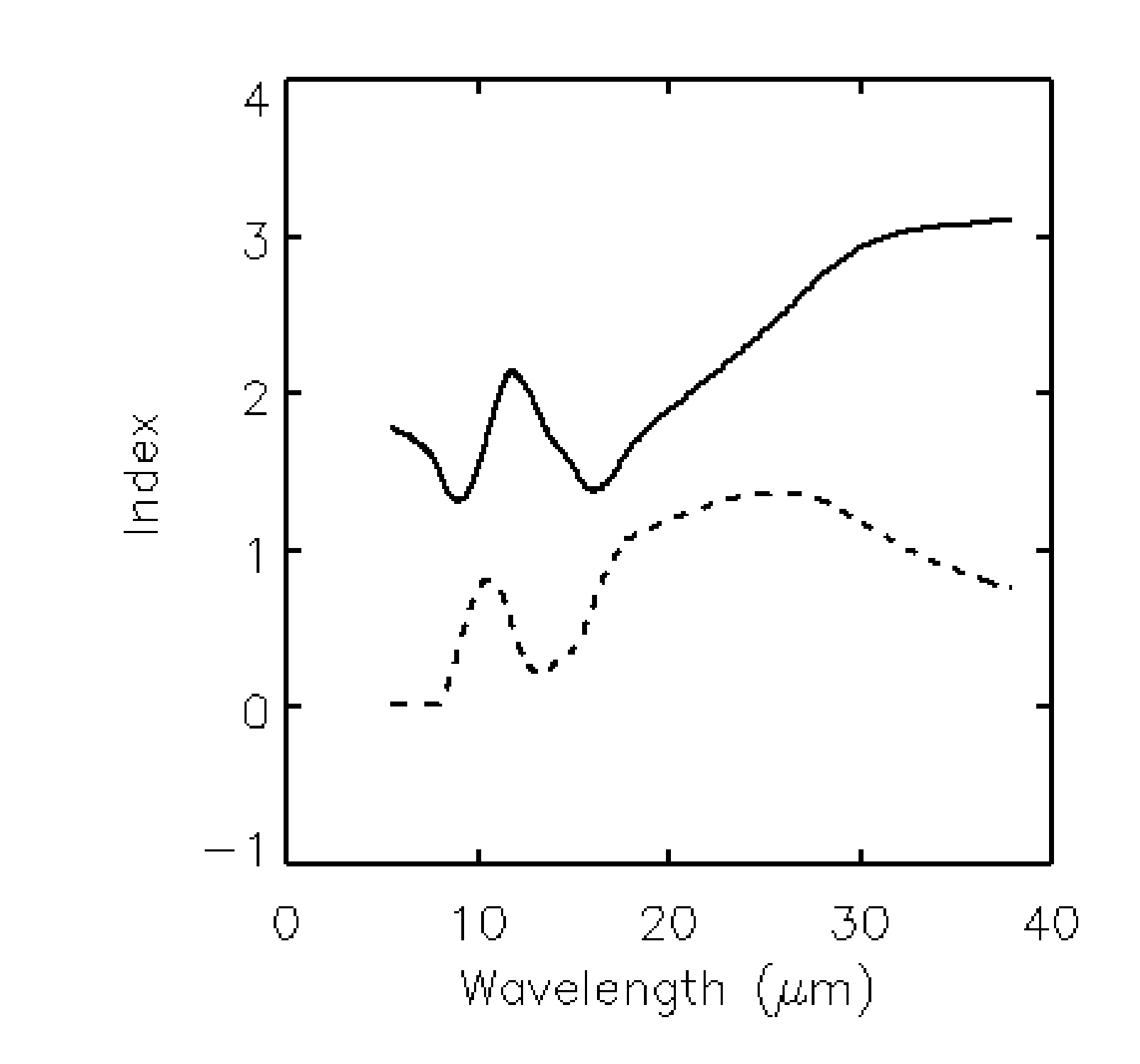}\\
\vspace{-0.3cm}\includegraphics[width=3in,angle=0]{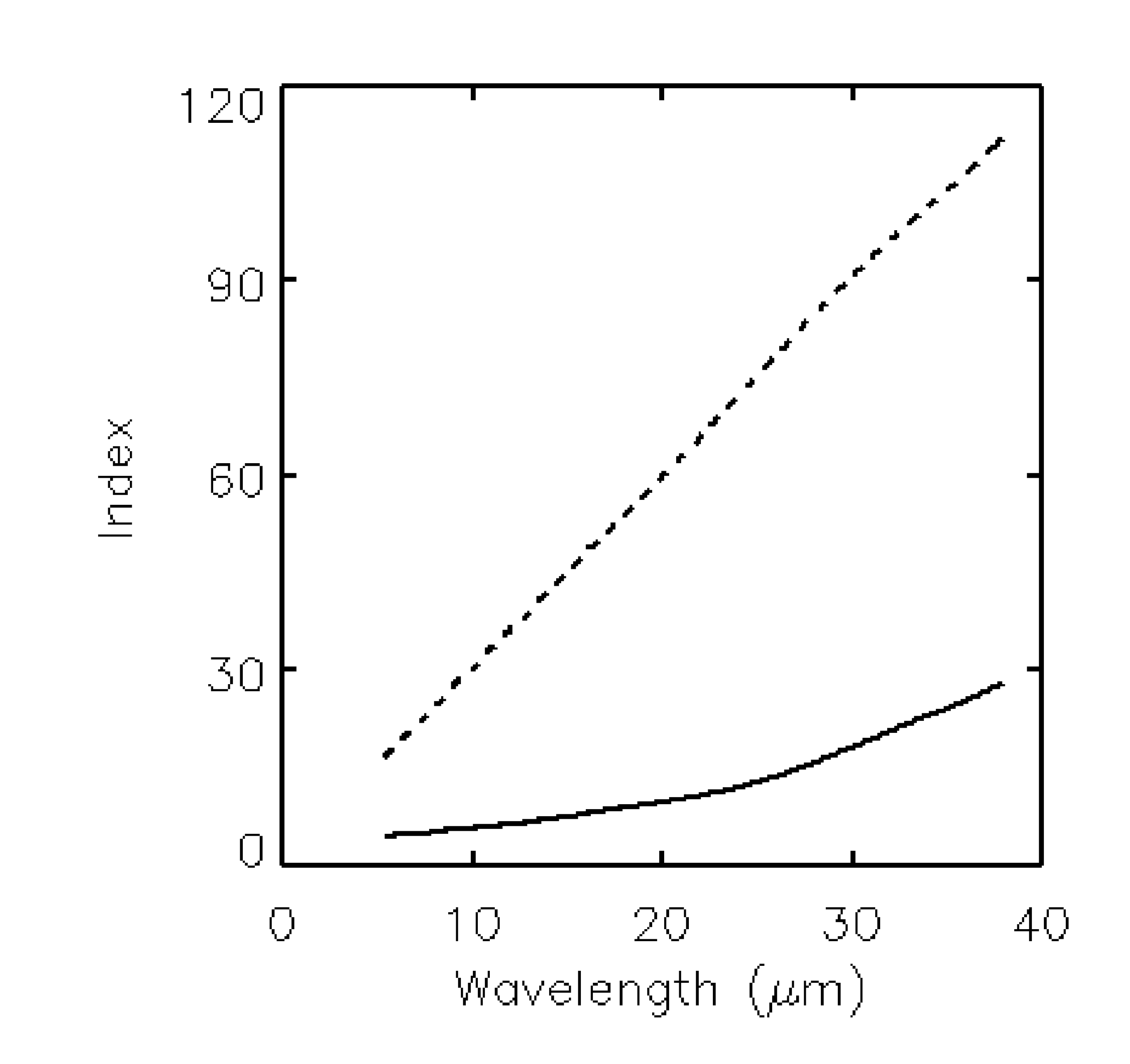}\\
\vspace{-0.3 cm}\includegraphics[width=3in,angle=0]{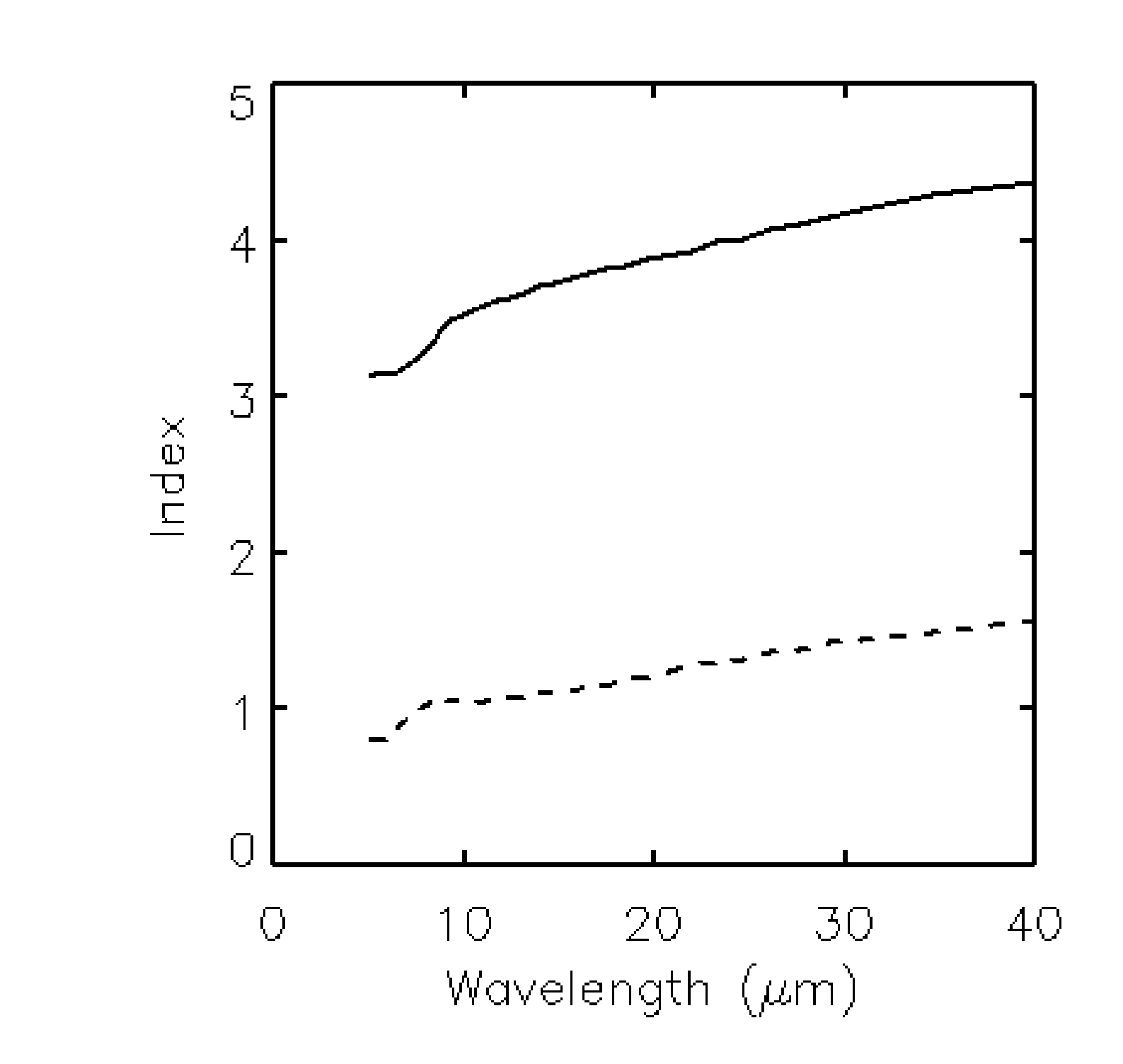}\\
\vspace{-0.6 cm}\caption{From top to bottom: A)  Optical constants of amorphous olivine from \citep{dorschner:1995} (dashed, k, solid, n).  The olivine is forsteritic with a composition of Fo80.  B) optical constants of iron from \citep{paquin:1995}; C) optical constants of carbon from \citep{preibisch:1993}.}
\label{fg1}
\end{center}
\end{figure}

We explore the effects of extremely fine-grained components and use models with a total of only three materials, of a palette of four:  A transparent matrix, assuming a real refractive index of 1.54 and an imaginary index of 0, representing many salts; two strong absorbers-- amorphous carbon and native iron--of micron size or less; and the silicate olivine. Olivine, carbon and iron are chosen as they are common constituents in cometary comae \citep{zolensky:2008, levasseur:2008} and asteroidal meteorites \citep{brown:2000, zolensky:2010}, and were detected in the Stardust sample collection. The optical constants for amorphous olivine (strictly speaking, a glass of olivine composition), carbon and iron used are shown in Figure \ref{fg1}. and are from \citet{dorschner:1995, paquin:1995, preibisch:1993}, respectively. In the model, the Trojan surface is represented by large grains of the transparent material (25 $\mu$m sizes are used throughout) with the small absorbers either infused throughout the transparent grain, or coating the surface (the effects of these two distributions are equivalent in this model). In any model we include only one of the two strong absorbers.  The models require more than one size of absorbing grain to match both the albedo and slope of the spectra. For simplicity, we used only two size grains, 1 micron grains, and grains that are a few tens of nanometers in size and have a Rayleigh size parameter $x$ less than 0.3 at all wavelengths. These extremely fine-sized particles have effects nearly independent of particle size \citep{hapke:2001, clark:2010a, clark:2010b, clark:2010c}. At thermal wavelengths both sizes have small $x$ parameters and only total abundance is relevant.

\subsection{Results}

\begin{figure}[b]
\begin{center}
\vspace{-1.0 cm}
\includegraphics[width=4.0in,angle=0]{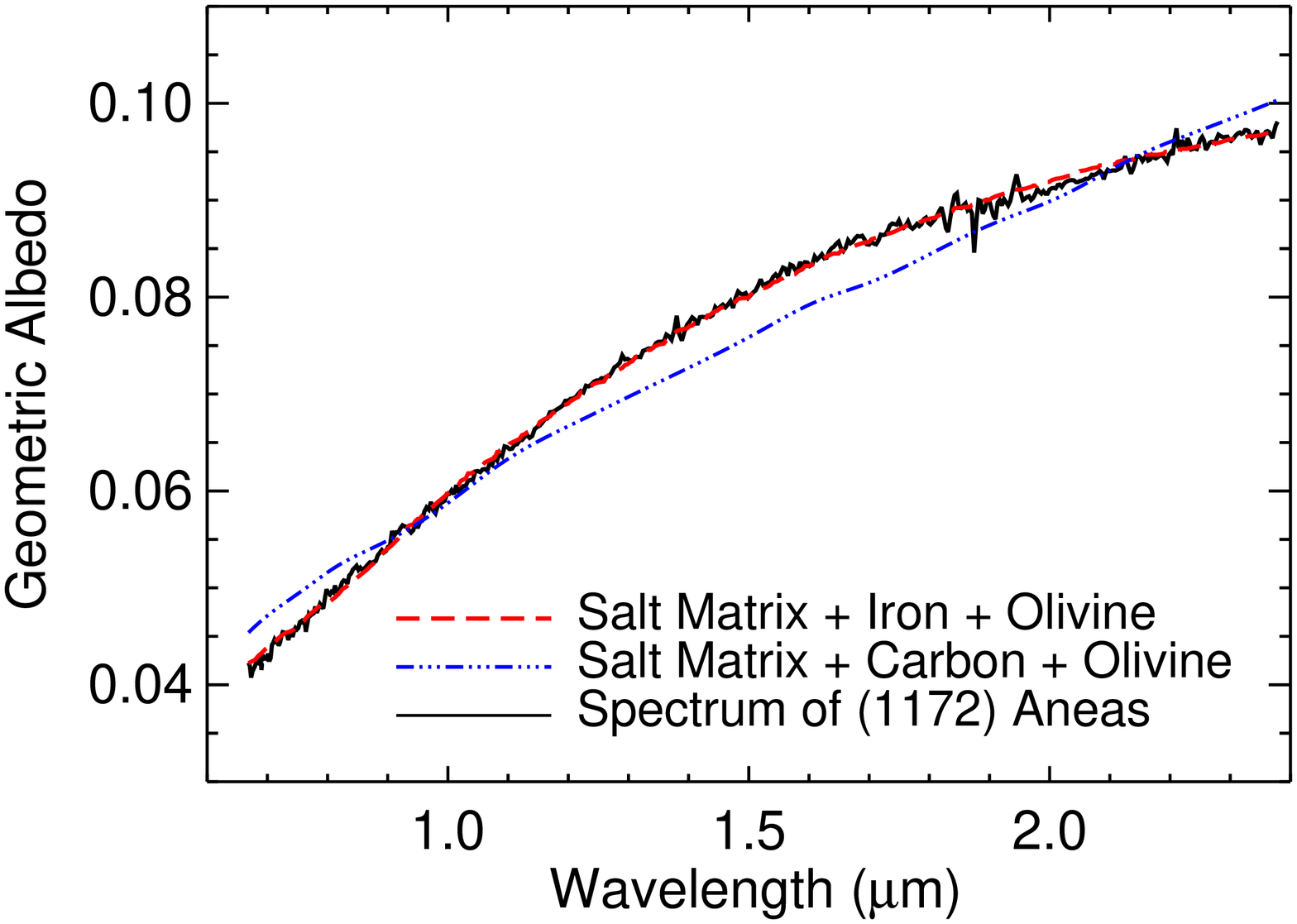}  \includegraphics[width=4.0in,angle=0]{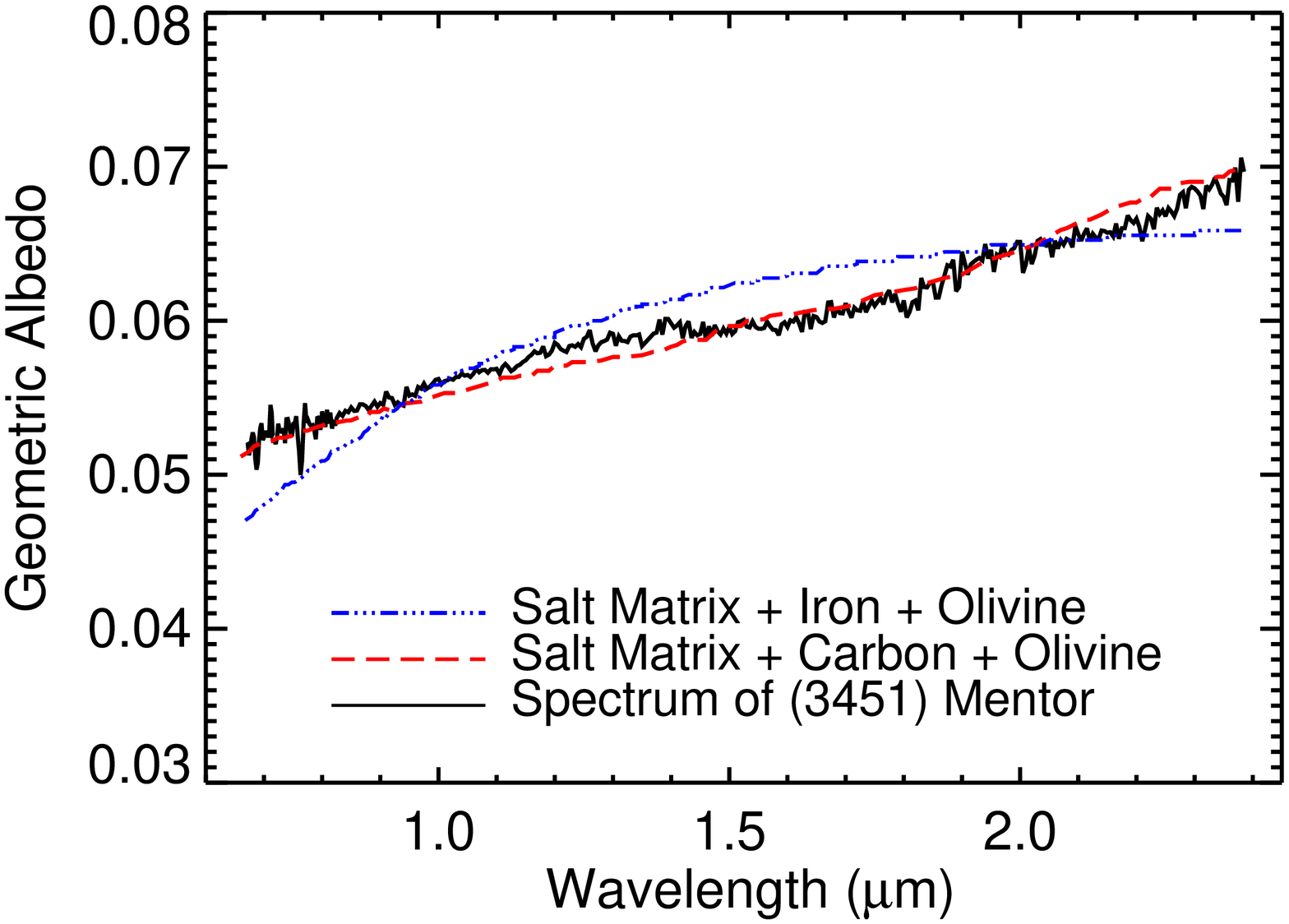}
\caption{$Upper$: a.) The solid line is the spectrum of a red Trojan (1172) Aneas taken with the IRTF telescope. The red dashed line is the modeled spectrum using 4 wt.\% $\mu$m-sized iron +  2 wt.\% nano-phase iron and 1 wt.\% of olivine suspended in a salt matrix. The blue dashed line is the modeled spectrum using 2 wt.\% $\mu$m-sized carbon + 14 wt.\% nano-phase carbon and 1 wt.\% of olivine suspended in a salt matrix.  $Lower$: b.) The solid line is the spectrum of a grey Trojan  (3451) Mentor, taken with the IRTF telescope.  The red dashed line is the modeled spectrum using 5 wt.\% $\mu$m-sized carbon +  6 wt.\% nano-phase carbon and 1 wt.\% of olivine suspended in a salt matrix. The blue dashed line is the modeled spectrum using 7 wt.\% $\mu$m-sized iron + 2 wt.\% nano-phase iron and 1 wt.\% of olivine suspended in a salt matrix. It shows that the iron model can best fit the red Trojan spectrum. in contrast, the spectrum of the grey Trojan can be explained by the carbon model. }
\label{fg2}
\end{center}
\end{figure}

The results of modeling the NIR spectra of both the ``red" and ``grey" Trojans are given in Figure \ref{fg2} and Table \ref{tab1}.  In the case of the ``red" Trojans, iron produced a better fit to the shape of the NIR spectrum (Figure \ref{fg2}a). The iron-bearing mixtures also require less mass of absorber than the carbon mixtures, with typically about 5 wt.\% and 15 wt.\% required respectively.  However, carbon cannot be ruled out entirely. Whereas, the ``grey" Trojans are better modeled with the carbon mixtures (Figure \ref{fg2}b). In the NIR, olivine is a very weak absorber at extremely small sizes, and principally acts as a scattering center, with an effect nearly identical to reducing the particle size of the transparent host grains. Therefore, the NIR spectra place only a weak constraint on olivine abundance. An upper limit of $\sim$5 wt.\% olivine is imposed to reproduce the low albedo of the asteroids, and best fits are achieved with 1 wt.\% olivine.  In summary, the NIR spectra of the Trojans are consistent with iron or carbon grains suspended in a transparent matrix with up to 5 wt.\% olivine permitted. 

Despite this success, we recognize that the abundances of absorber required to match the albedo of the Trojan spectra might erase the very silicate feature that is the crucial clue to the transparent matrix. To investigate this possibility, we used the same model at thermal wavelengths, and acquired spectra of physical mixtures (described in the next section).  A significant challenge to modeling in this region is the presence of thermal gradients in the optical surface that have a very strong effect on spectra in the region of thermal emission.  There have been a variety of attempts to model this effect, but even models of simple reflectance are only modestly successful at reproducing measured spectra in detail \citep[e.g.\ ][]{mustard:1997}. Nevertheless, these are the tools at hand.  Specifically, we use the Lucey and Riner$'$s model (2011) applied at thermal wavelengths.  We note that Hapke-based theories, which relate optical constants to single scattering albedo, are not applicable in regions of strong absorption, for example in the case of silicates. Given that salts are not strong absorbers, our adoption of the Hapke model is therefore valid.

\begin{figure}[h]
\begin{center}
\vspace{0.5 cm}
\includegraphics[width=5.5in,angle=0]{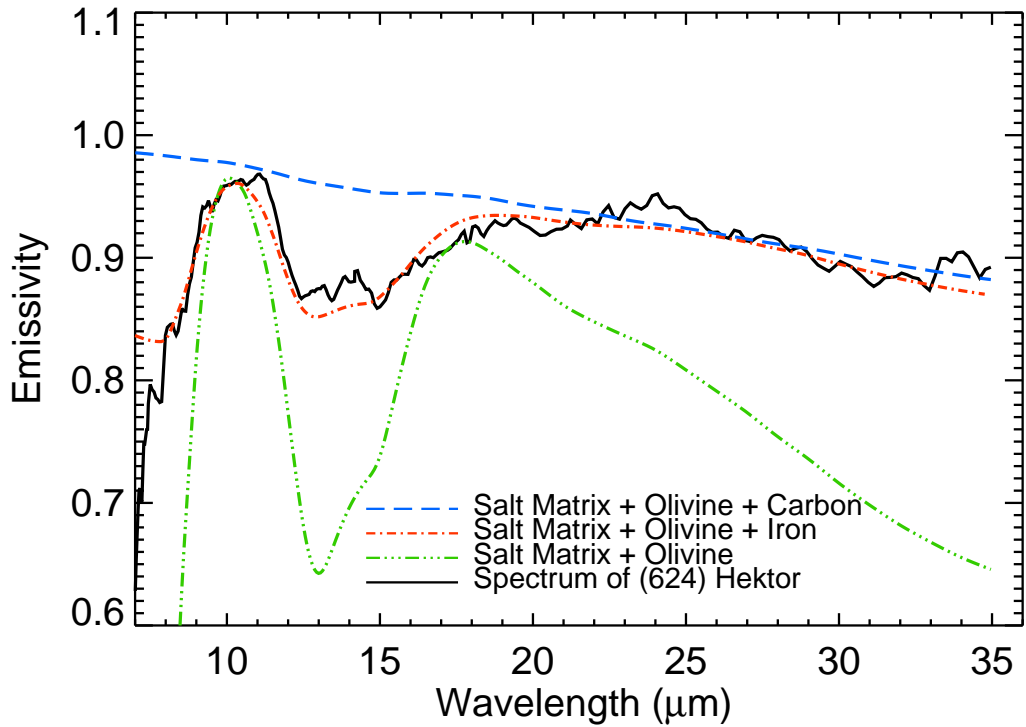}
\caption{The solid line is the spectrum of ``red" Trojan (624) Hektor, taken with the Spitzer Space telescope and is provided by Dr. Josh Emery. The red dashed line is the modeled spectrum using 5 wt.\% of iron and 5 wt.\% of olivine suspended in a salt matrix (25\% salt + 65\% neutral absorber). The blue dashed line is the modeled spectrum using 10 wt.\% of carbon and 1 wt.\% of olivine suspended in a salt matrix. The model that uses only olivine in a salt matrix is shown in green.}
\label{fg3}
\end{center}
\end{figure}

\begin{figure}[h]
\begin{center}
\vspace{0.5 cm}
\includegraphics[width=5.5in,angle=0]{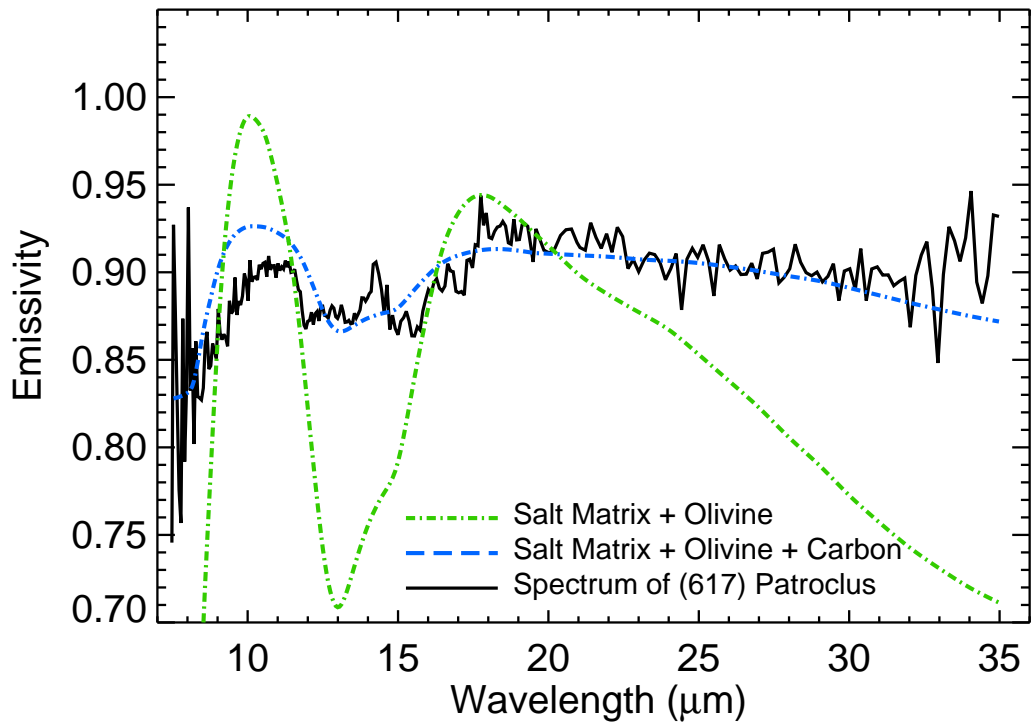}
\caption{.  The solid line is the spectrum of the ``grey" Trojan (617) Patroclus, taken with the Spitzer Space telescope by \cite{mueller:2010} and is provided by Dr.\ Josh Emery. The blue dashed line is the modeled spectrum using 5 wt.\% of carbon and 5 wt.\% of olivine suspended in a salt matrix. The model that uses only olivine in a salt matrix is shown in green. It illustrates that carbon sufficiently suppresses the 10 $\mu$m emission feature. Our carbon model fits the grey Trojan spectrum adequately well. }
\label{fg4}
\end{center}
\end{figure}

Applying the abundances arrived at in the NIR model exercise to thermal wavelengths, we find that the iron mixtures and carbon mixtures have distinct effects on the silicate features in the TIR. The iron model (shown as the red dashed line in Figure \ref{fg3}), though does not match the Trojan spectrum in detail (no fine-tuning the model to fit individual spectral features has been attempted and our use of olivine alone is simplistic), the key 10 $\mu$m silicate feature is well preserved in the modeled TIR spectrum. In contrast, the silicates features are more severely impacted by a carbon contaminant as shown in blue in Figure \ref{fg3}.  On the other hand, the carbon model matches the sample spectrum of ``grey" Trojans adequately well in the TIR, see Figure \ref{fg4}, which is consistent with our finding in the NIR, 

It has been noted that only ``red" Trojans, such as (624) Hektor, show a prominent 10 $\mu$m emission feature in the TIR. Whereas, the ``grey" Trojans, such as (617) Patroclus, show none or much weaker 10 $\mu$m emission in their TIR spectra \citep{emery:2008}. Our modeling results are consistent with the previous Spitzter observations of Trojans and suggest that the ``red" Trojans contain nano-phased iron as the strong absorber which absorbs a large fraction of visible light and, in turn, causes the low albedo and the red spectral slope. Since the silicate emission feature is not affected by iron at longer wavelengths, the ``red" Trojans tend to exhibit strong emission feature near 10 $\mu$m. The ``grey" Trojans, on the other hand, have been contaminated with fine-grained carbon particles and the silicate feature is severely muted by the carbon component. Given that carbon is not as efficient as iron in absorbing UV light, it explains the less red spectral slopes of the ``grey" Trojans. 

\section{Laboratory Experiments}
To better understand the potential weaknesses in the thermal spectral model, we acquired a series of laboratory MIR emissivity spectra of olivine-salt-carbon mixtures. Like our model, these data were not acquired under vacuum with simulated solar heating that is known to have a striking effect on thermal emission spectra, strengthening contrast in spectra of very fine particles, but we use these data to inform us regarding the visibility of the silicate feature. We acquired all of the spectra on a Nicolet 6700 FTIR spectrometer. Emissivity measurements between 2000 and 250 cm$^{-1}$ (5-40 $\mu$m) were carried out using a CsI beamsplitter and a deuterated triglycine sulfate (DTGS) detector with a CsI window. Samples were heated to 80 $^{\circ}$C to increase signal, and spectra were calibrated using a blackbody source according to the methods of \citet{ruff:1997}. It should be noted that the calibration of radiance data to emissivity assumes unit emissivity somewhere in the wavelength region of interest. Because of the relatively low emissivity of halite between 5 and 40 $\mu$m, this may not actually be the case, and a slope may be imparted on the spectrum \cite[e.g.]{osterloo:2008}. 
	
Our mixtures were composed of two or three components, which included halite, San Carlos olivine, and amorphous carbon. Nanophase iron was not used as a component owing to its rapid alteration to iron oxide. San Carlos olivine was ground in an agate ball mill and the $<$ 2 $\mu$m size fraction was separated according to Stokes$^{\prime}$ settling method at 22 $^{\circ}$C using HPLC grade distilled water. We used a reagent grade 99\%+ pure synthetic halite powder for our salt phase and submicron amorphous carbon for the strong absorbing phase. We prepared three mixtures using these starting products: 1) salt + 1 wt.\% carbon, 2) salt + 2 wt.\% olivine, and 3) salt + 1 wt.\% carbon + 2 wt.\% olivine. After mixing, samples were homogenized by gently grinding with an agate mortar and pestle.

\begin{figure}[b]
\begin{center}
\vspace{0.5 cm}
\includegraphics[width=5in,angle=0]{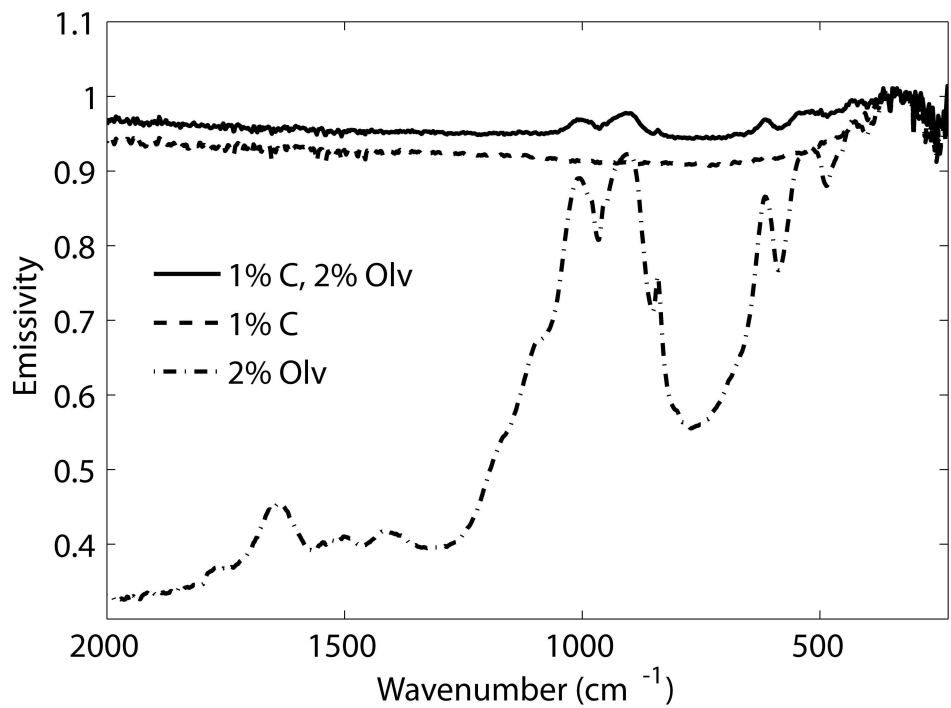}
\caption{Spectra of mixtures of salt, olivine and carbon.  In this case the olivine is crystalline and has more spectral structure than the amorphous olivine used in the modeling.}
\label{fg5}
\end{center}
\end{figure}

MIR emissivity spectra of the three mixtures are displayed in Figure \ref{fg5}. The salt/carbon mixture is essentially featureless with emissivity values less than unity over much of the spectral range (features between 30 and 40 $\mu$m in all three spectra are due to minor water vapor in the atmosphere during measurement). In contrast, the salt/olivine mixture displays strong features throughout the spectrum, with major absorptions centered at 8 and 13 $\mu$m. When 1\% C is added to the salt/olivine mixture, the prominent silicate absorption features are significantly weakened, but emissivity maxima at $\sim$10 and 11 $\mu$m still persist. Our laboratory measurements is entirely consistent with our TIR model and demonstrate that carbon is indeed effective at suppressing the silicate feature. However , the weakening effect due to carbon has not been reported or discussed in the literature concerning simulations of comet spectra in the TIR. In these studies, large amount (20-30 wt.\%) of amorphous carbon are often used \citep{wooden:2000, min:2005}. The model and the laboratory data reveal a novel effect of a combination of a non-absorbing matrix (salt) and minute amounts of a strong absorber (carbon) on typical silicate emissivity. 

\section{Discussion}

While the proposition by \citet{emery:2006} that an unusual scattering situation on the measured Trojans may be required by the thermal data does not rule out some future, more prosaic explanations, this is the state-of-art in understanding the surfaces of these objects.  Of the two suggestions by Emery et al., a fairy castle structure or a transparent medium, the former is known to occur (on the Moon at least, Hapke 1968) as a natural consequence of micrometeorite impact or vacuum deposition processes.  However, it has not been demonstrated that a fairy castle structure can quantitatively (or qualitatively) account for the spectral properties of the Trojans observed in the thermal IR. The latter hypothesis, that silicates are suspended in a transparent medium, has wider implications for the evolution of the Trojans. Based on their laboratory measurements \citet{king:2011} specifically suggested that salt could be that medium on the Trojans. Our work here shows that a transparent medium can account for the NIR spectral properties and still preserve a 10 $\mu$m silicate feature. Thus, a weakly contaminated transparent medium is the only scattering scenario that can explain observations for both wavelength regions, although high surface porosity has also been suggested to explain the TIR (but not the NIR) spectra \citep{vernazza:2012}. Local salt rich surfaces are observed on the Earth, Mars \citep{osterloo:2008} and Europa \citep{mccord:2001}. Prior to superior data from Voyager, salt is suggested for Io \citep{fanale:1974}. Also, brine volcanism is proposed as an important geological process for the icy satellites of the outer planets \citep{kargel:1991}. However, on the Earth and Mars these surfaces are highly restricted and are geologic exceptions; in the Jovian system the satellites enjoy especially vigorous heating which is not a likely part of the history of the Trojans. Our study begs the question whether a salt-rich surface is plausible for the Trojan asteroids. 

One possible origin for a salt-rich surface is deposition as an evaporite layer, and we explore the plausibility of this hypothesis for the Trojans. We begin with a discussion about whether water ice could be present in the Trojans. Provided the Trojans originated beyond the snow line, high ice abundances are expected \citep{jewitt:2004}, though Jewitt et al.\ point out that no observational evidence supports this assumption. But similar studies have made that assumption for objects lie within 5 AU.  For example, \citet{mccord:2005} suggested that Ceres contains about 25\% ice (and incidentally, proposed salts on its surface to explain its high albedo). The discovery of the main belt comets, that probably formed in the asteroid belt at about 3 AU \citep{hsieh:2006}, underscores the presence of ice at heliocentric distances well within the orbit of Jupiter.  Meteoritical evidence is also relevant.  Many characteristics of carbonaceous chondrites testify to high early water abundances within their parent bodies including veins of salt, low temperature aqueous alteration minerals and the requirement for a more efficient cooling mechanism than conduction to prevent high temperatures in the face of Al$^{26}$ heating (see MSween et al. 2002 for a summary). Although the knowledge of the initial ice-to-rock-ratio of the Trojans is  poorly constrained, first-order estimate of possible composition can be yielded via density. Both (624) Hektor and (617) Patroclus are binary systems and their densities are measured to be 2.48 g/cc \citep{lacerda:2007} and 0.8 g/cc \citep{marchis:2006}, respectively. These two large objects represent a range of possible ice-fraction in the Trojans: from 4 wt.\%, assuming the silicate density as low as that of serpentine [2.54 g/cc], up to nearly 100 wt.\%. 

In the ice-rich scenario (shown as Figure 6a), assuming a primordial Trojan accreted early with respect to Al$^{26}$ decay, short-lived radionuclides promote internal melting of the ice and the silicate alteration could go to completion with liquid water remaining. The volume expansion accompanying alteration and evolution of volatiles such as carbon dioxide would lead to high internal pressures, driving residual liquid water toward the surface. In a carbonaceous chondrite parent body there are abundant volatiles and mechanisms to provide pressure to drive volcanism.  Ammonia \citep{mccord:2005} and carbon dioxide \citep{mcsween:2002} are likely companion volatiles to ice, and as the ice melts vapor is produced.  Carbon dioxide and hydrogen may be produced through alteration reactions \citep{wilson:1999,mcsween:2002}.  Gas produced from alteration was shown by Wilson et al.\ to provide enough energy to completely disrupt some asteroids. For 100 km-sized objects, such as the large Trojans, available energy is probably sufficient to propel several percent of their liquid contents toward the surface of the asteroid. In contact with the altered silicates, the water would take up soluble ions and become a brine. For example, \citet{fanale:1974} exposed a sample of the carbonaceous chondrite Orgueil to ``near boiling" water for one hour and produced a brine containing a mass of soluble ions equal to 5 wt.\% of the mass of the original meteorite sample, and at about 20 wt.\% in the brine, was substantially below saturation. Isotopic evidence suggests in the case of the carbonaceous chondrite parents the water temperatures were lower, but the exposure times must have been far longer, even for ``one pass" exposure (see Young et al. 1999).

We note that the actual composition of the rock component of Trojans is currently unclear, however, it could take up one of the two possible forms. One possibility is that the rock component resembled CI or CM carbonaceous chondrites, then Trojans would have contained initially abundant water-soluble substances. Alternatively, the rock portion had never experienced aqueous processes prior to or during accretion, then Trojans would have consisted of ice intermixed with anhydrous silicates resembling C3 or ordinary chondrites. In this scenario, as the heating source melted the internal ice, which was in close contact with rock, aqueous reactions between liquid water and anhydrous silicates would be inevitable, yielding hydrated phyllosilicates and soluble salts. Thus, regardless of the initial alteration status of the rock component, partial or complete melting of the water ice probably would lead to the formation of brines rather than pure water. 

Any early asteroid is clad with a cold primitive thermal crust that would cause freezing of upwelling liquid water under it, forming a possibly global sill of frozen brine a few kilometers below the surface. However, if the initial ice-rock ratio is too low (shown as Figure 6a), no water remains after silicate alteration.  In this case the asteroid, no longer buffered by liquid water, will continue to heat, eventually dehydrating the silicates, and again providing a source of liquid water to take up ions, and deposit a brine below the chill crust.

In either scenario the asteroid is exposed to the intense early impact flux. Trojans are well confined at two Lagrangian points and have a wider distribution in their inclinations than that of the main belt asteroids. Therefore the impact rate for Trojan asteroids is higher than the inner solar system \citep{delloro:1998,melita:2009}. We take the model developed for Ida presented in \citep{geissler:1994} and apply it to the Trojan population to estimate the erosional rate.  Assuming a flux of 6.5$\times$10$^{-18}$ km$^{-2}$ yrs$^{-1}$ and characteristic impact velocity of 5 km/s \citep{melita:2009}, we find that around 3$\times$10$^9$ kg/yr of material will be stripped from the surface during excavation of craters.This corresponds to a shell about 20-km thick every billion years. These estimates are based on canonical scaling laws for vertical impacts, however, and therefore should be considered as upper limits.  A true model taking into account the obliquity and updated velocity scaling laws promises a more realistic estimate.  

As discussed above, intense bombardments are likely to erode the primitive crust and may expose the frozen brine layer. The exposed brine would promptly sublimate until a protective armor of salt accumulates. Following the late heavy bombardment era, occasional impact events may re-expose frozen brine which would sublimate and again armor with salt. Hydrated minerals may be present on the Trojans and could be exposed by impacts, however, their spectral signatures may be masked by the subsequent formation of an evaporite layer.  

\section{Conclusions}
In this paper we show that the hypothesis by \citep{emery:2006} that the surface of the Trojans may be composed of silicate grains in a transparent matrix is consistent with the NIR spectra of the Trojans. We also show through theory and experiment that the required abundances of carbon as the dark absorber (in the case of the ``gray" Trojans) significantly weakens the silicate emission band in the TIR. On the other hand, we show that iron (in the case of the ``red" Trojans) appears not to erase the silicate features giving rise to the original transparent matrix hypothesis. We expand upon the suggestion by \citep{king:2011} that the transparent matrix is salt, and explicitly suggest an evaporite surface for the Trojans that results from sublimation of a deep layer of frozen brine exposed by impacts. Prior work has shown that hydrothermal circulation has occurred within carbonaceous chondrite parent bodies -- reasonable (although not the best) analogies to the early Trojans. Previous studies have shown that substantial quantities of brine can be produced as a consequence of hydrothermal circulation and have identified a number of pressure sources with sufficient magnitude to drive brine volcanism towards the surface. With time, the brine will freeze under the cold primitive crust.  Subsequently, the thermal crust is removed by impacts, exposing frozen brine. By sublimation, salt is concentrated in the optical surface. Fine-grained silicates, carbon, and iron are common products of sublimation of comets and asteroid collisions. In particular, carbon and iron are found in abundance in Stardust samples from comet Wild 2 and are expected contaminants on outer solar system objects. Over the Solar System's history, salt layers are contaminated and colored by fine dust from comets or impact remnants. 

\section{Acknowledgment}
We thank Dr.\ Josh Emery for providing the original Spitzer spectrum of Trojans Hektor and Patroclus, Dr.\ Brendan Hermalyn for the impact modeling on the Trojans and Dr.\ David Jewitt for commenting on the manuscript. We would like to thank the referees, Dr.\ Josh Emery and Dr.\ Roger Clark for the constructive comments that substantially helped improving the manuscript. BY was supported by the National Aeronautics and Space Administration through the NASA Astrobiology Institute under Cooperative Agreement No. NNA08DA77A issued through the Office of Space Science. TG would like to acknowledge the Stony Brook College of Arts and Sciences.


\clearpage

\begin{thebibliography}{34}

\bibitem[Brown et al.(2000)]{brown:2000}Brown, P.~G., Hildebrand, 
A.~R., Zolensky, M.~E., et al.\ 2000, Science, 290, 320 

\bibitem[Christensen(1978)]{christensen:1978} Christensen, N.\ 1978, 
Tectonophysics, 47, 131 

\bibitem[Clark et al.(2010a)]{clark:2010a} Clark, R.~N., Cruikshank, 
D.~P., Jaumann, R., et al.\ 2010, Bulletin of the American Astronomical 
Society, 42, 952 

\bibitem[Clark et al.(2010b)]{clark:2010b} Clark, R.~N., Pieters, 
C.~M., Taylor, L.~A., et al.\ 2010, Lunar and Planetary Institute Science 
Conference Abstracts, 41, 2337 

\bibitem[Clark et al.(2010c)]{clark:2010c}
Clark, R.~N.,  {\it et~al.\/}, AGU Fall Meeting, abstract, P13E, 3 (2010). 

\bibitem[{Cruikshank {et~al.}(2001)Cruikshank, Dalle~Ore, Roush, Geballe, Owen,
  de~Bergh, Cash, \& Hartmann}]{cruikshank:2001}
Cruikshank, D.~P., Dalle~Ore, C.~M., Roush, T.~L., Geballe, T.~R., Owen, T.~C.,
  de~Bergh, C., Cash, M.~D., \& Hartmann, W.~K. 2001, Icarus, 153, 348
  
  \bibitem[dell'Oro et 
al.(1998)]{delloro:1998} dell'Oro, A., PAolicchi, F., Marzari, P., Dotto, E., \& Vanzani, V.\ 1998, A\&A, 339, 272 


\bibitem[{Dorschner {et~al.}(1995)Dorschner, Begemann, Henning, Jaeger, \&
  Mutschke}]{dorschner:1995}
Dorschner, J., Begemann, B., Henning, T., Jaeger, C., \& Mutschke, H. 1995,
  A\&A, 300, 503

\bibitem[{Dotto {et~al.}(2006)Dotto, Fornasier, Barucci, Licandro, Boehnhardt,
  Hainaut, Marzari, de~Bergh, \& de~Luise}]{dotto:2006}
Dotto, E., {et~al.} 2006, Icarus, 183, 420

\bibitem[{Emery {et~al.}(2011)Emery, Burr, \& Cruikshank}]{emery:2011}
Emery, J.~P., Burr, D.~M., \& Cruikshank, D.~P. 2011,  AJ,
  141, 25

\bibitem[{Emery(2008)}]{emery:2008}
Emery, J.~P. 2008, in New Horizons in Astronomy: Frank N. Bash Symposium 2007
  ASP Conference Series, ed. A.~F. J. R. M.~J. Shen; \& M.~H. Siegel, Vol. 393,
  San Francisco: Astronomical Society of the Pacific, 3

\bibitem[{Emery {et~al.}(2006)Emery, Cruikshank, \& van Cleve}]{emery:2006}
Emery, J.~P., Cruikshank, D.~P., \& van Cleve, J. 2006, Icarus, 182, 496

\bibitem[{Emery \& Brown(2004)}]{emery:2004}
Emery, J.~P., \& Brown, R.~H. 2004, Icarus, 170, 131

\bibitem[{Emery \& Brown(2003)}]{emery:2003}
Emery, J.~P., \& Brown, R.~H. 2003, Icarus, 164, 104

\bibitem[{Fanale et~al.(1974)Fanale, Johnson, and Matson}]{fanale:1974}
Fanale, F.~P., Johnson, T.~V., Matson, D.~L., 12 1974, Science 186, 922.

\bibitem[{Fern\`andez {et~al.}(2003)Fern\`andez, Sheppard, \&
  Jewitt}]{fernandez:2003}
Fern\`andez, Y.~R., Sheppard, S.~S., \& Jewitt, D.~C. 2003, 
AJ, 126, 1563

\bibitem[{Fleming \& Hamilton(2000)}]{fleming:2000}
Fleming, H.~J., \& Hamilton, D.~P. 2000, Icarus, 148, 479

\bibitem[{Fornasier {et~al.}(2007)Fornasier, Dotto, Hainaut, Marzari,
  Boehnhardt, de~Luise, \& Barucci}]{fornasier:2007}
Fornasier, S., Dotto, E., Hainaut, O., Marzari, F., Boehnhardt, H., de~Luise,
  F., \& Barucci, M.~A. 2007, Icarus, 190, 622

\bibitem[Geissler et al.(1994)]{geissler:1994} Geissler, P., Petit, 
J.-M., \& Greenberg, R.\ 1994, Lunar and Planetary Institute Science Conference Abstracts, 25, 411 

\bibitem[{Grav {et~al.}(2011)Grav, Mainzer, Bauer, Masiero, Spahr, McMillan,
  Walker, Cutri, Wright, Eisenhardt, Blauvelt, DeBaun, Elsbury, Gautier,
  Gomillion, Hand, \& Wilkins}]{grav:2011}
Grav, T., {et~al.} 2011, ApJ, 742

\bibitem[Hapke(1968)]{hapke:1968} Hapke, B.\ 1968, Science, 159, 
76 

\bibitem[{Hapke(1981)}]{hapke:1981}
Hapke, B., 1981, JGR, 86, 3039

\bibitem[{Hapke(1993)}]{hapke:1993}
Hapke, B., 1993, Theory of reflectance and emittance spectroscopy. Topics in Remote Sensing, Cambridge, UK: Cambridge University Press

\bibitem[{Hapke(2001)}]{hapke:2001}
Hapke, B., 2001, JGR, 106, 10039

\bibitem[{Hsieh \& Jewitt(2006)}]{hsieh:2006}
Hsieh, H.~H., \& Jewitt, D. 2006, Science, 312, 561

\bibitem[{Jewitt {et~al.}(2004)Jewitt, Sheppard, \& Porco}]{jewitt:2004}
Jewitt, D.~C., Sheppard, S., \& Porco, C. 2004, Jupiter's outer satellites and
  Trojans, In: Jupiter. The planet, satellites and magnetosphere, Fran Bagenal, Timothy E. Dowling, William B. McKinnon (eds), 
  Cambridge, UK: Cambridge University Press, P. 263--280

\bibitem[{Jewitt {et~al.}(2000)Jewitt, Trujillo, \& Luu}]{jewitt:2000}
Jewitt, D.~C., Trujillo, C.~A., \& Luu, J.~X. 2000, AJ, 120,
  1140

\bibitem[Kargel(1991)]{kargel:1991} Kargel, J.~S.\ 1991, Icarus, 
94, 368 

\bibitem[{King {et~al.}(2011)King, Izawa, Vernazza, McCutcheon, Berger, \&
  Dunn}]{king:2011}
King, P.~L., Izawa, M. R.~M., Vernazza, P., McCutcheon, W.~A., Berger, J.~A.,
  \& Dunn, T. 2011, Lunar and Planetary Institute Science Conference Abstracts, 42, 1985 

\bibitem[{Lacerda \& Jewitt(2007)}]{lacerda:2007}
Lacerda, P., \& Jewitt, D.~C. 2007, AJ, 133

\bibitem[{Levasseur et~al.(2008)}]{levasseur:2008}A.~C. Levasseur-Regourd, Zolensky, M., \& Lasue, J.\, (2008). P\&SS, 56, 1719.

\bibitem[{Lucey \& Noble(2008)}]{lucey:2008}
Lucey, P.~G., \& Noble, S.~K. 2008, Icarus, 197, 348

\bibitem[{Lucey \& Riner(2011)}]{lucey:2011}
Lucey, P.~G., \& Riner, M.~A. 2011, Icarus, 212, 451

\bibitem[Macke et 
al.(2011)]{macke:2011} Macke, R.~J., Consolmagno, G.~J., \& Britt, D.~T.\ 2011, Meteoritics and Planetary Science, 46, 1842 

\bibitem[Marchis et al.(2006)]{marchis:2006} Marchis, F., 
Hestroffer, D., Descamps, P., et al.\ 2006, Nature, 439, 565 

\bibitem[{Marzari \& Scholl(1998)}]{marzari:1998}
Marzari, F., \& Scholl, H. 1998, Icarus, 131, 41

\bibitem[{McCord and Sotin(2005)}]{mccord:2005}
McCord, T.~B., Sotin, C., 2005, JGR, 110, E05009

\bibitem[{McCord {et~al.}(2001)McCord, Orlando, Teeter, Hansen, Sieger, Petrik,
  \& Van~Keulen}]{mccord:2001}
McCord, T.~B., Orlando, T.~M., Teeter, G., Hansen, G.~B., Sieger, M.~T.,
  Petrik, N.~G., \& Van~Keulen, L. 2001, JGR, 106,
  3311

\bibitem[McSween et al.(2002)]{mcsween:2002} McSween, H.~Y., Jr., 
Ghosh, A., Grimm, R.~E., Wilson, L., \& Young, E.~D.\ 2002, in Asteroids III, W. F. Bottke Jr., A. Cellino, P. Paolicchi, and R. P. Binzel (eds),
 University of Arizona Press, Tucson, p.559-571 

\bibitem[Melita et al.(2009)]{melita:2009} Melita, M.~D., 
Strazzulla, G., \& Bar-Nun, A.\ 2009, Icarus, 203, 134 


\bibitem[{Min {et~al.}(2005)Min, Hovenier, de~Koter, Waters, \&
  Dominik}]{min:2005}
Min, M., Hovenier, J.~W., de~Koter, A., Waters, L. B. F.~M., \& Dominik, C.
  2005, Icarus, 179, 158

\bibitem[{Morbidelli {et~al.}(2005)Morbidelli, Levison, Tsiganis, \&
  Gomes}]{morbidelli:2005}
Morbidelli, A., Levison, H.~F., Tsiganis, K., \& Gomes, R. 2005, Nature, 435,
  462

\bibitem[Mueller et al.(2010)]{mueller:2010} Mueller, M., Marchis, 
F., Emery, J.~P., et al.\ 2010, Icarus, 205, 505 

\bibitem[Mustard \& Hays(1997)]{mustard:1997} Mustard, J.~F., \& Hays, J.~E.\ 1997, Icarus, 125, 145 

\bibitem[{Osterloo {et~al.}(2008)Osterloo, Hamilton, Bandfield, Glotch,
  Baldridge, Christensen, Tornabene, \& Anderson}]{osterloo:2008}
Osterloo, M.~M., et al. 2008, Science, 319, 1651

 \bibitem[{Paquin(1995)}] {paquin:1995}R. A.~Paquin, Properties of Metals, in Handbook of Optics Volume II: Devices, Measurements, and Properties, edited by M. Bass, E. W. Van Stryland, D. R. Williams and W. L. Wolfe, pp. 35.31-35.78, McGraw-Hill, Inc., New York (1995).

\bibitem[Postberg et al.(2006)]{postberg:2006} Postberg, F., Kempf, 
S., Srama, R., et al.\ 2006, Icarus, 183, 122 

\bibitem[{Preibisch {et~al.}(1993)Preibisch, Ossenkopf, Yorke, \&
  Henning}]{preibisch:1993}
Preibisch, T., Ossenkopf, V., Yorke, H.~W., \& Henning, T. 1993, A\&A, 279, 577

\bibitem[{Ruff {et~al.}(1997)Ruff, Christensen, Barbera, \&
  Anderson}]{ruff:1997}
Ruff, S.~W., Christensen, P.~R., Barbera, P.~W., \& Anderson, D.~L. 1997,
  JGR, 102, 14899

\bibitem[Vernazza et al.(2012)]{vernazza:2012} Vernazza, P., Delbo, 
M., King, P.~L., et al.\ 2012, LPI Contributions, 1667, 6049 


\bibitem[{Wilson et~al.(1999)Wilson, Keil, Browning, Krot, and
  Bourcier}]{wilson:1999}
Wilson, L., Keil, K., Browning, L.~B., Krot, A.~N., \& Bourcier, W.\ 1999, Meteoritics and Planetary Science, 34, 541

\bibitem[{Wooden {et~al.}(2000)Wooden, Butner, Harker, \&
  Woodward}]{wooden:2000}
Wooden, D.~H., Butner, H.~M., Harker, D.~E., \& Woodward, C.~E. 2000, Icarus,
  143, 126

\bibitem[{Yang \& Jewitt(2011)}]{yang:2011}
Yang, B., \& Jewitt, D. 2011,  AJ, 141, 95

\bibitem[{Yoshida \& Nakamura(2005)}]{yoshida:2005}
Yoshida, F., \& Nakamura, T. 2005, AJ, 130, 2900

\bibitem[{Zolensky et~al.(2010)}] {zolensky:2010} Zolensky, M., et al.\, 2010. Meteoritics and Planetary Science, 45, 1618.

\bibitem[{Zolensky {et~al.}(2008)Zolensky, Nakamura-Messenger, Rietmeijer,
  Leroux, Mikouchi, Ohsumi, Simon, Grossman, Stephan, Weisberg, Velbel, Zega,
  Stroud, Tomeoka, Ohnishi, Tomioka, Nakamura, Matrajt, Joswiak, Brownlee,
  Langenhorst, Krot, Kearsley, Ishii, Graham, Dai, Chi, Bradley, Hagiya,
  Gounelle, \& Bridges}]{zolensky:2008}
Zolensky, M., {et~al.} 2008, Meteoritics and Planetary Science, 43, 261
\end{thebibliography}

 \clearpage

\begin{table}[hb]
\hspace{-3.in}\caption{Model parameters for NIR Trojan Spectra}
\begin{tabular}{cccccccc}
\hline
\hline
Trojan & Absorber & Salt Matrix & 1$\mu$m Absorber & Nano. Absorber & Olivine \\       
   &                     wt.\%  &   wt.\%  &   wt.\% &   wt.\% &   wt.\%\\
\hline             
Aneas & Iron &  93.0 & 4.0 & 2.0 & 1.0\\
Aneas & Carbon &83.0 & 2.0 & 14.0 & 1.0 \\
Mentor & Iron &  90.0 & 7.0 & 2.0 & 1.0\\
Mentor& Carbon &88.0 & 5.0 & 6.0 & 1.0 \\
\hline
\hline
\end{tabular}
\label{tab1}
\end{table}

\begin{table}[!ht]
\hspace{-3.in} \caption{Model parameters for TIR Trojan Spectra}
\begin{tabular}{cccccccc}
\hline
\hline
Trojan & Salt Matrix &  Absorber & Olivine \\       
   &             wt.\%  &   wt.\%  &   wt.\% \\
\hline             
Hektor &  25.0 + 65.0 & 5.0 (Iron) & 5.0 \\
Hektor &  89.0 & 10.0 (Carbon) & 1.0 \\
Hektor & 99.0 & - & 1.0 \\
Patroclus & 90.0 & 5.0 (Carbon) & 5.0 \\
Patroclus & 99.0 & - & 1.0 \\
\hline
\hline
\end{tabular}
\label{tab2}
\end{table}

\vspace{-3.in}

\begin{figure}[b]
\vspace{-7 cm}
\includegraphics[width=5in,angle=0]{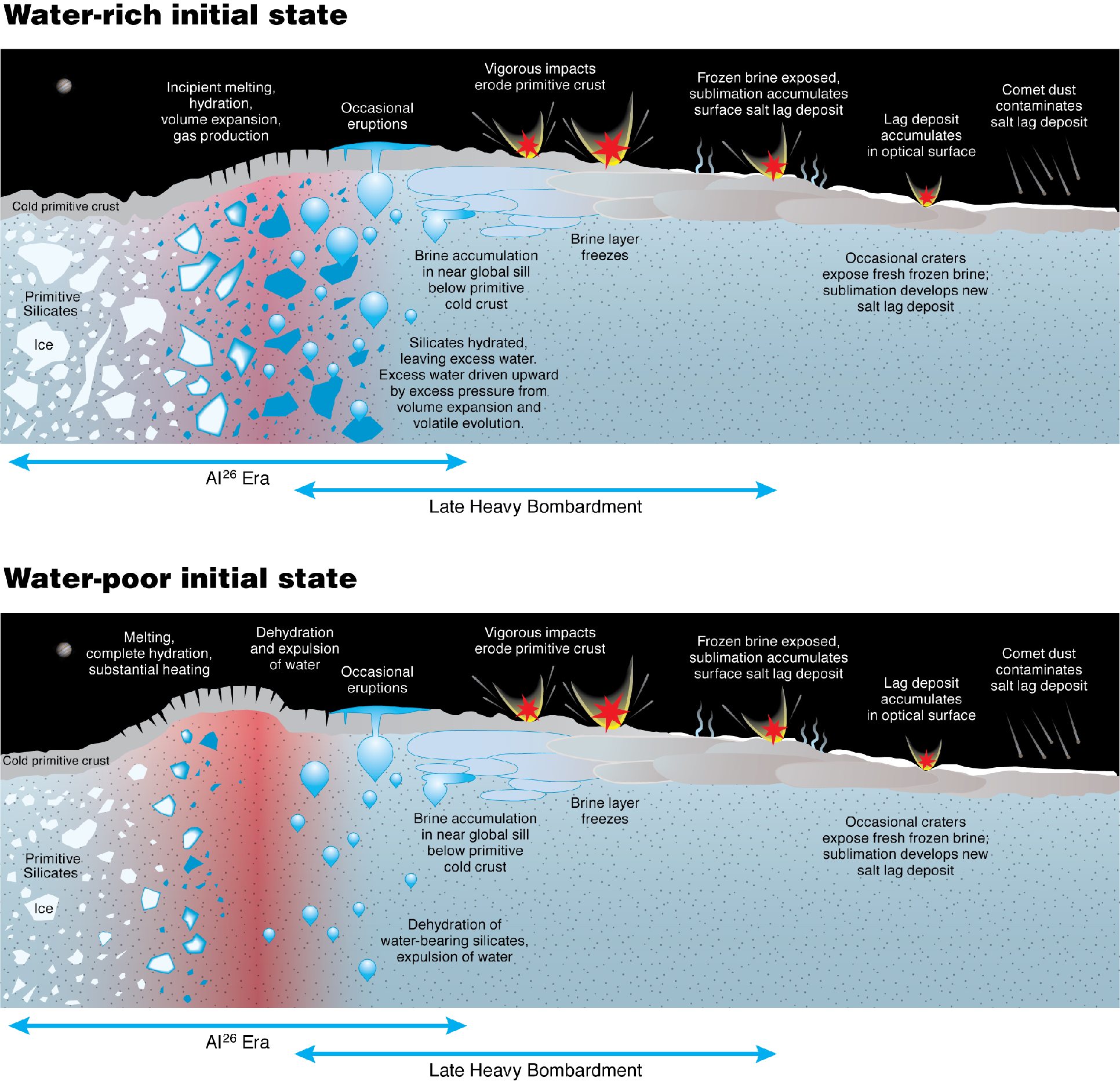}
\caption{Schematic of the internal and surface evolution of Trojan-like primitive objects. On assembly, the object is composed of a mechanical mixture primitive silicates and ices. As Al$^{26}$ decays the ice melts and silicates begin to alter. Depending on the initial ice content the object can follow two paths.  In the ice-rich case (top; a) the silicates are completely hydrated, leaving residual liquid water. During hydration and melting of ices, gas species that are more volatile than water build up the pressure in the interior of the object, forcing circulating liquid water upward where it will tend to accumulate below the cold primitive surface layer.  The liquid water, a brine because of contact with the rock component, will eventually freeze below the primitive crust.  Simultaneously, the late heavy bombardment will erode the primitive crust and may expose the frozen brine layer.  The exposed brine will promptly sublimate until a protective armor of salt accumulates.  Following the late heavy bombardment, occasional impact events may re-expose frozen brine which will sublimate and again armor with salt.  The ice-poor initial state (bottom; b) follows a different early evolution, but similar late evolution.  In this scenario, ice is completely consumed by the altering silicates, and without the buffering effect of melting and alteration the interior can heat substantially until the silicates dehydrate, again contributing liquid water to the system, opportunities for dissolution of ions, and accumulation in a sub-crustal layer.}
\label{fg6}
\end{figure}


\end{document}